\documentclass[twocolumn,prd,unsortedaddress,superscriptaddress,showpacs,a4paper,nofootinbib]{revtex4-1} 
\usepackage{epsfig}
\usepackage{subfigure} 
\usepackage{amsmath} 
\usepackage[dvips]{color}

\def\beq{\begin{equation}} 
\def\enq{\end{equation}} 
\def\beqa{\begin{eqnarray}} 
\def\enqa{\end{eqnarray}}

\def\qq{\lag\bar{q}q\rag}

\def\qGq{\lag\bar{q}Gq\rag}

\def\gGG{\lag g_s^2 G^2 \rag}

\def\ga{\gamma}

\def\lb{\label} 
 
\def\nn{\nonumber}

\newcommand{\rag}{\rangle} 
\newcommand{\lag}{\langle}

\def\MeV{\mbox{ MeV}} 
\def\GeV{\mbox{ GeV}} 

\begin{document} 
 
\title{\sc Production of the $Y(4260)$ state in B meson decay} 

\author{R.M. Albuquerque}
\email{raphael@ift.unesp.br}
\affiliation{Instituto de F\'{i}sica Te\'{o}rica, Universidade Estadual Paulista (IFT-UNESP)\\  
R.\,Dr.\,Bento\,Teobaldo\,Ferraz,  271 - Bl.\,II Sala 207, 01140-070 S\~ao Paulo/SP - Brasil}
\author{M.~Nielsen} 
\email{mnielsen@if.usp.br}   \affiliation{Instituto   de  F\'{\i}sica, 
  Universidade de S\~{a}o Paulo, C.P.  66318, 05389-970 S\~{a}o Paulo, 
  SP,    Brazil} 
\author{C.M. Zanetti}
\email{carina.zanetti@gmail.com}
\affiliation{ Faculdade de Tecnologia, Universidade do Estado do Rio de 
Janeiro, Rod. Presidente Dutra Km 298, P\'olo Industrial, 27537-000, 
Resende/RJ - Brasil}
 
\begin{abstract} 
 
We calculate the branching ratio for the production of the meson $Y(4260)$ 
in the decay $B^- \to Y(4260)K^-$. We use QCD sum rules approach and we 
consider the $Y(4260)$ to be a mixture between charmonium and exotic 
tetraquark, $[\bar{c}\bar{q}][qc]$, states with $J^{PC}=1^{--}$. Using the value 
of the mixing angle determined previously as: $\theta=(53.0\pm0.5)^\circ$, we 
get the branching ratio 
 $\mathcal{B}(B\to Y(4260)K)=(1.34\pm0.47)\times10^{-6}$, which allows us to 
 estimate an interval on the branching fraction 
 $3.0 \times 10^{-8} < {\mathcal B}_{_Y} < 1.8 \times 10^{-6}$ in agreement with 
 the experimental upper limit reported by Babar Collaboration.
 
\end{abstract}  
 
\maketitle 
 
 
The $Y(4260)$ state was first observed by BaBar collaboration in the 
$e^+e^-$ annihilation through initial state radiation \cite{babar1}, and it 
was confirmed by CLEO and Belle collaborations \cite{yexp}. The $Y(4260)$ was 
also observed in the $B^-\to Y(4260)K^-\to J/\Psi\pi^+\pi^-K^-$ decay 
\cite{babary2}, and CLEO reported two additional decay channels: 
$J/\Psi\pi^0\pi^0$ and $J/\Psi K^+K^-$ \cite{yexp}. 
The $Y(4260)$ is one of the many charmonium-like state, called  $X,~Y$ and 
$Z$ states, recently observed in  $e^+e^-$ collisions 
by BaBar and Belle collaborations that do not fit the quarkonia interpretation.
The production mechanism, masses, decay widths, spin-parity 
assignments and decay modes of these states have been discussed in some 
reviews  \cite{Zhu:2007wz,Nielsen:2009uh,Brambilla:2010cs,Nielsen:2014mva}.
The $Y(4260)$ is particularly interesting because some new states have
been identified in the decay channels of the $Y(4260)$, like the 
$Z_c^+(3900)$. The $Z_c^+(3900)$ was first observed by the BESIII 
collaboration in the $(\pi^\pm J/\psi)$ mass spectrum 
of the $Y(4260)\to J/\psi\pi^+\pi^-$ decay channel \cite{Ablikim:2013mio}.
This structure, was also observed at the same time by the  
Belle collaboration \cite{Liu:2013dau} and was confirmed by the  authors of 
Ref. ~\cite{Xiao:2013iha} using  CLEO-c data.  
 
The decay  modes of the  $Y(4260)$ into $J/\psi$ and  other charmonium 
states indicate the existence of a $\bar{c}c$ in its content.  However, 
the attempts  to classify this  state in the   charmonium spectrum 
have failed since the $\Psi(3S),~\Psi(2D)$ and $\Psi(4S)$ $c\bar{c}$ states 
have been assigned to the well established $\Psi(4040),~\Psi(4160),~$ and 
$\Psi(4415)$ mesons respectively, and the prediction from quark models 
for the $\Psi(3D)$  state is 4.52 GeV. Therefore, the mass of the $Y(4260)$
is not consistent with any of the $1^{--}$ $c\bar{c}$ states 
\cite{Zhu:2007wz,Nielsen:2009uh}. 

Some theoretical interpretations for the $Y(4260)$ are: 
tetraquark state \cite{tetraquark}, hadronic $D_{1} D$, $D_{0} D^*$ molecule 
\cite{Ding}, $\chi_{c1} \omega$ molecule \cite{Yuan}, $\chi_{c1} \rho$ molecule
 \cite{liu}, $J/\psi f_0(980)$ molecule \cite{oset}, 
a hybrid charmonium \cite{zhu}, a charm-baryonium \cite{Qiao}, a cusp 
\cite{eef1,eef2,eef3}, etc. Within the available experimental information, 
none of these suggestions can be completely ruled out. However, there are some
calculations, within the QCD sum rules (QCDSR) approach
\cite{Nielsen:2009uh,svz,rry,SNB}, that can not explain 
the mass of the $Y(4260)$ supposing it to be a tetraquark state \cite{rapha}, 
or  a $D_{1} D$, $D_{0} D^*$ hadronic molecule \cite{rapha}, or  a  
$J/\psi f_0(980)$ molecular state \cite{Albuquerque:2011ix}. 

In the framework of the QCDSR  the mass and the decay width, 
in the channel  $J/\psi\pi\pi$, of the $Y(4260)$ were 
computed  with  good   agreement  with  data,  considering it as a 
mixing between two and four-quark states \cite{Dias:2012ek}.
The mixing is done at the level of the hadronic currents and, physically, this  
corresponds to a fluctuation of the $c \overline{c}$ state where a gluon is 
emitted and subsequently splits into a light quark-antiquark pair, which 
lives for some time and behaves like a tetraquark-like state. The same
approach was applied to the $X(3872)$ state and good agreement with the data
were obtained for its mass and the  decay width  into $J/\psi\pi\pi$ 
\cite{x3872mix}, its radiative decay \cite{x3872rad}, and also in the 
$X(3872)$ production rate in $B$ decay \cite{x3872prod}. 

In this work we will focus on the  production of the $Y(4260)$, using the 
mixed two-quark and four-quark prescription of Ref.~\cite{Dias:2012ek} 
to perform a QCDSR analysis of the process $B^-\to Y(4260)K^-$.
The experimental upper limit on the branching fraction 
for such a production in $B$ meson decay has been reported by BaBar 
Collaboration \cite{babary2}, with $95\%$ C.L.,  
\begin{equation} 
  \label{branching} 
	{\mathcal B}_{_Y} <\! 2.9\times10^{-5} 
\end{equation} 
where ${\mathcal B}_{_Y} \equiv {\mathcal B}
(B^- \!\!\to\! K^- Y(4260),Y(4260) \!\to\! J/\psi\pi^+\pi^-)$.
 
 
\begin{figure}[h,t,b]
\centerline{ 
\includegraphics[width=0.35\textwidth]{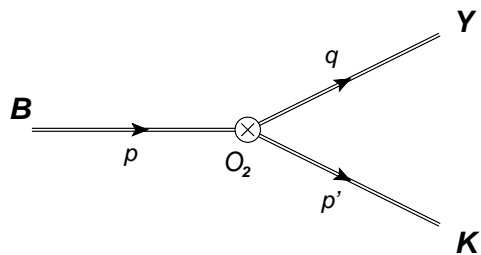}
}
\caption{The process for production of the $Y(4260)$ state in B meson decay, 
mediated by an effective vertex operator ${\mathcal O}_2$.}
\label{weakdecay}
\end{figure}

The process  $B\to Y(4260)K$ occurs via  weak decay of  the $b$ quark, while 
the $u$ quark  is a spectator. The $Y$ meson 
as a mixed  state of tetraquark and charmonium  interacts via $\bar{c}c$ 
component  of the  weak current. In effective theory, at  the scale 
$\mu\sim m_b\ll m_W$, the weak  decay is treated as a four-quark local 
interaction described by the effective Hamiltonian (see Fig.~\ref{weakdecay}): 
\begin{equation}\lb{ham} 
  {\mathcal{H}}_W=\frac{G_F}{\sqrt{2}}V_{cb}V_{cs}^*\left[\left(C_2(\mu)+ 
\frac{C_1(\mu)}{3}\right) 
    {\mathcal{O}}_2+\cdots\right]\,, 
\end{equation} 
where $V_{ik}$ are CKM  matrix elements, $C_1(\mu)$ and $C_2(\mu)$ are 
short  distance  Wilson coefficients  computed  at the  renormalization 
scale $\mu\sim{\mathcal O}(m_b)$.   The four-quark effective operator 
is ${\mathcal{O}}_2=J_{\mu}^{(\bar{c}c)}J_{\mu}^W$, with
\begin{equation}\lb{wcurrents}    J_{\mu}^W=\bar{s}\Gamma_\mu   b\,,\quad 
  J_{\mu}^{(\bar{c}c)}=\bar{c}\Gamma_\mu  c\,,\end{equation}
and
$\Gamma_\mu=\gamma_\mu(1-\gamma_5)$. 
 
Using factorization, 
the decay amplitude of the process is calculated from the Hamiltonian  
(\ref{ham}), by splitting the matrix element in two pieces: 
\begin{eqnarray}\lb{amp} 
{\mathcal M} 
&=&i\frac{G_F}{\sqrt{2}}V_{cb}V_{cs}^*\left(C_2+\frac{C_1}{3}\right)\nn\\& 
\times&\langle B(p)\vert J_{\mu}^W\vert K(p^\prime)\rangle\langle Y(q) 
\vert J^{\mu(\bar{c}c)}\vert0\rangle, 
\end{eqnarray} 
where      $p=p^\prime+q$.
Following Ref.~\cite{x3872prod}, the matrix
elements in Eq.~({\ref{amp}}) are parametrized as:
\begin{equation}\lb{2pmatrix} 
\langle Y(q)\vert J_{\mu}^{(\bar{c}c)}\vert0\rangle=\lambda_W 
\epsilon^\ast_\mu(q)\,, 
\end{equation} 
and 
\begin{equation}\lb{3pmatrix} 
\langle B(p)\vert J_{\mu}^W\vert K(p^\prime)\rangle=f_+(q^2)(p_\mu+ 
p_\mu^\prime)+f_-(q^2)(p_\mu-p_\mu^\prime)\,. 
\end{equation} 
The  parameter  $\lambda_W$  in  (\ref{2pmatrix}) gives  the  coupling 
between the current $J_\mu^{(\bar{c}c)}$  and the $Y$ state.  The form 
factors $f_\pm(q^2)$  describe the weak transition $B\to  K$. Hence we 
can see  that the  factorization of the  matrix element  describes the 
decay as two separated sub-processes. 
 
The decay width for the process $B^-\to Y(4260)K^-$ is given by  
\begin{equation}\lb{eqwidth} 
\Gamma(B\to YK)=\frac{\vert{\mathcal{M}}\vert^2}{16\pi m_B^3}\sqrt{\lambda(m_B^2,m_K^2,m_Y^2)}, 
\end{equation} 
with $\lambda(x,y,z)=x^2+y^2+z^2-2xy-2xz-2yz$. The invariant amplitude 
squared can be obtained from (\ref{amp}), using (\ref{2pmatrix})
and (\ref{3pmatrix}): 
\begin{eqnarray} 
\vert\mathcal{M}\vert^2&=&\frac{G_F^2}{2 m_Y^2}\vert V_{cb}V_{cs}\vert^2\left(C_2 
+\frac{C_1}{3}\right)^2\nn\\\nn\\ 
&\times& \lambda(m_B^2,m_K^2,m_Y^2)\lambda_W^2f_+^2
\,. 
\end{eqnarray} 

The coupling constant $f_ +$ was determined in Ref.\cite{x3872prod} through 
extrapolation of the form factor $f_+(Q^2)$ to the meson pole $Q^2 = - m_Y^2$, 
using the QCDSR approach for the three-point correlator \cite{bcnn}: 
\beqa 
\Pi_{\mu}(p,p^\prime)&=&\!\!\int d^4x \,d^4y \,e^{i(p^\prime\cdot x-\,p 
\cdot y)}\langle0\vert T\{J_\mu^W(0)\times\nn\\
&\times&J_K(x)J^\dagger_B(y)\}\vert0\rangle, 
\enqa  
where the  weak  current, $J^W_\mu$, is  defined  in 
(\ref{wcurrents})  and  the interpolating  currents  of  the  $B$ and  $K$ 
pseudoscalar mesons are: 
\begin{equation} 
  J_K = i\,\bar{u}_a \gamma_5 s_a \,,\quad J_B=i\,\bar{u}_a 
\gamma_u b_a\,. 
\end{equation}  

The obtained result for the form factor was \cite{x3872prod}:
\begin{equation}\lb{fplus} 
  f_+(Q^2)=\frac{(17.55\pm0.04) \GeV^2}{(105.0\pm1.8)\GeV^2+Q^2}\,. 
\end{equation} 

For the decay width calculation, we  need the value of the form factor 
at  $Q^2=-m_Y^2$,  where  $m_Y$  is  the mass  of the $Y(4260)$ meson. 
Using $m_Y=(4251\pm9)\MeV$ \cite{pdg} we get: 
\begin{equation}\lb{fpluspolo} 
  f_+(Q^2)\vert_{Q^2=-m_Y^2}=0.206\pm0.004\,. 
\end{equation} 

The  parameter $\lambda_W$ can also be determined using the QCDSR approach
for the two-point  correlator: 
\begin{equation} 
\Pi_{\mu\nu}(q)=i\int d^4y~e^{iq\cdot y}\langle0\vert T\{J_\mu^Y(y) 
J_\nu^{(\bar{c}c)}(0)\}\vert0\rangle\,, 
\label{2point}
\end{equation} 
where    the    current    $J_\nu^{(\bar{c}c)}$    is    defined    in 
(\ref{wcurrents}).  For the $Y$  meson we will follow \cite{Dias:2012ek} and 
consider    a    mixed     charmonium-tetraquark current: 
\beq
J_\mu^Y = \sin \theta \:J_\mu^{(4)} + \cos\theta \:J_\mu^{(2)},
\label{jmix}
\enq
where
\beqa
J_\mu^{(4)} &=& \frac{\epsilon_{abc} \epsilon_{dec}}{\sqrt{2}}
\Big[(q_a^T C\ga_5 c_b)(\bar{q}_d \ga_\mu\ga_5 C\bar{c}_e^T)+\nn\\
&&~+(q_a^T C\ga_5\ga_\mu c_b)(\bar{q}_d\ga_5 C\bar{c}_e^T) \Big],
\label{j4q}
\enqa
\beq
J_\mu^{(2)}=\frac{1}{\sqrt{2}}\qq \:\Big( 
\bar{c}_a\ga_\mu c_a \Big) ~\equiv~ \frac{1}{\sqrt{2}}\qq \:J_\mu^{'(2)} ~.
\enq
In Eq.~(\ref{jmix}), $\theta$ is the mixing angle that was determined in
\cite{Dias:2012ek} to be: $\theta=(53.0\pm0.5)^0$.

Inserting the currents (\ref{wcurrents}) 
and (\ref{jmix}) in the correlator we have in the OPE side of the sum 
rule 
\begin{eqnarray} 
\Pi^{\mathrm{OPE}}_{\mu\nu}(q)&=&\sin\theta\, 
\Pi^{4,2}_{\mu\nu}(q)
+\frac{\qq}{\sqrt{2}}\cos\theta\, 
\Pi^{2,2}_{\mu\nu}(q)\,, 
\end{eqnarray} 
where 
\begin{eqnarray} 
\Pi^{4,2}_{\mu\nu}(q)&=&i\int d^4y ~e^{iq\cdot y}\langle0\vert T\{ 
J_\mu^{(4)}(y)J_{\nu(\bar{c}c)}(0)\}\vert0\rangle\nn\\ 
\Pi^{2,2}_{\mu\nu}(q)&=&i\int d^4y ~e^{iq\cdot y}\langle0\vert T\{ 
J_\mu^{'(2)}(y)J_{\nu(\bar{c}c)}(0)\}\vert0\rangle\,. 
\label{pi24}
\end{eqnarray} 
Only the vector   part    of   the   current $J_\nu^{(\bar{c}c)}$  contributes
to the correlators in Eq.~(\ref{pi24}). Therefore, these  correlators are  the 
same as the ones calculated  in Ref.~\cite{Dias:2012ek} for  the mass of the 
$Y(4260)$. 
 
\begin{table}[t]
\setlength{\tabcolsep}{1.25pc}
\caption{\small QCD input parameters.}
\begin{tabular}{ll}
&\\
\hline
Parameters&Values\\
\hline
$\overline{m}_c$ & $(1.23 - 1.47) \GeV$ \\
$\qq$ & $ \hspace{-0.25cm}-(0.23 \pm 0.03)^3\GeV^3$\\
$\gGG$ & $(0.88 \pm 0.25)~\GeV^4$\\
$m_0^2 \equiv \qGq/\qq$ & $(0.8 \pm 0.1) ~\GeV^2$\\
\hline
\end{tabular}
\label{Param}
\end{table}

To evaluate the  phenomenological side we insert intermediate states of the 
$Y$: 
\begin{eqnarray} 
  \Pi_{\mu\nu}^{phen}(q)&=&\frac{i}{q^2-m_Y^2}\langle0\vert J^Y_\mu\vert  
Y(q)\rangle\langle Y(q)\vert J^{(\bar{c}c)}_\nu\vert0\rangle\,,\nn\\ 
&=&\frac{i\lambda_Y\lambda_W}{Q^2+m_Y^2}\left(g_{\mu\nu}-\frac{q_\mu q_\nu} 
{m_Y^2}\right) 
\end{eqnarray} 
where $q^2=-Q^2$,  and we  have used the  definition (\ref{2pmatrix}) 
and 
\begin{equation} 
  \langle0\vert J^Y_\mu\vert Y(q)\rangle=\lambda_Y\epsilon_\mu(q)\,. 
\end{equation} 
The parameter $\lambda_Y$, that defines the  coupling between the current 
$J^Y_\mu$ and the $Y$ meson, was determined in Ref.~\cite{Dias:2012ek}
to be: $\lambda_Y = (2.00 \pm 0.23) \times 10^{-2} ~ \GeV^5$. 

As usual in the QCDSR approach, we perform a Borel  transform to $Q^2\to M_B^2$ 
to  improve the matching between both  sides of the sum  rules. 
After performing the Borel transform in both sides of the sum rule we  get in 
the $g_{\mu\nu}$ structure: 
\begin{eqnarray}\lb{2psumrule} 
  \lambda_W\lambda_Ye^{-\frac{m_Y^2}{M_B^2}}= 
\frac{\sin \theta}{\sqrt{2}}\,\Pi^{4,2}(M_B^2)
+\frac{\qq}{\sqrt{2}}  \cos \theta\, 
\Pi^{2,2}(M_B^2) ~~~~~~
\end{eqnarray} 
where the invariant functions $\Pi^{2,2}(M_B^2)$ and $\Pi^{4,2}(M_B^2)$ are
written in terms of a dispersion relation, 
\begin{eqnarray}
   \Pi(M_B^2) = \int\limits_{4m_c^2}^{\infty} \!ds ~e^{-s/M_B^2} \:\rho(s) ~~,
\end{eqnarray}
with their respective spectral densities $\rho^{2,2}(s)$ and $\rho^{4,2}(s)$
given in Appendix.

\begin{figure}[bp]
\begin{center}
\includegraphics[width=0.45\textwidth]{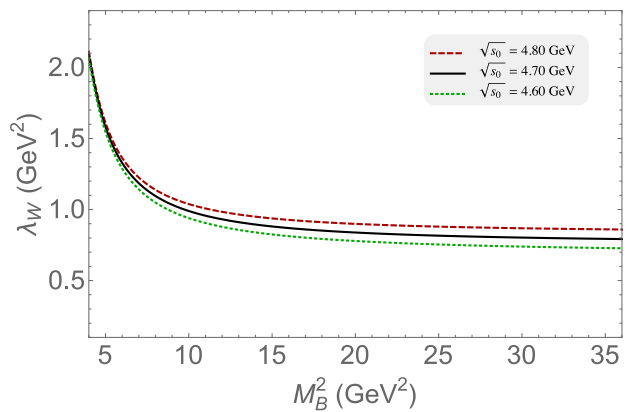}
\caption{The coupling parameter $\lambda_W$ as a function of $M_B^2$, for 
different values of the continuum threshold.}
\label{figLW}
\end{center}
\end{figure}

We  perform  the calculation of the  coupling parameter  $\lambda_W$  
using the  same values for  the masses and QCD condensates as in 
Ref.~\cite{Dias:2012ek} which are listed in Table \ref{Param}. To be 
consistent with the calculation of $\lambda_Y$ we also use the same 
region in the threshold  parameter $s_0$ as in Ref.~\cite{Dias:2012ek}: 
$\sqrt{s_{0}} = (4.70 \pm 0.10)$ GeV. As one can see in Fig.~\ref{figLW}, 
the region where we get $M_B^2$-stability is given by: 
$(8.0 \leq M_B^2 \leq 25.0) \GeV^2$.  

Taking into account the variation in the Borel mass parameter, in the
continuum threshold, in the quark condensate, in the coupling constant $\lambda_Y$ 
and in the mixing angle $\theta$, the result for the $\lambda_W$ parameter is: 
\begin{equation}\lb{lambdaW} 
  \lambda_W=(0.90\pm0.32)\GeV^2\,. 
\end{equation} 

Thus we can calculate the decay width in Eq.~(\ref{eqwidth}) by using the values of 
$f_+(-M_Y^2)$ and $\lambda_W$, determined in Eqs.~(\ref{fpluspolo}) and 
(\ref{lambdaW}). The  branching   ratio  is evaluated 
dividing   the  result   by  the   total  width   of  the   $B$  meson 
$\Gamma_{\mathrm{tot}}=4.280 \times 10^{-4} \:\mbox{eV}$: 
\begin{equation}\lb{result} 
  \mathcal{B}(B\to Y(4260)K)=(1.34\pm0.47)\times10^{-6}\,, 
\end{equation} 
where   we   have   used  the  CKM parameters  $V_{cs}=1.023$, 
$V_{cb}=40.6\times10^{-3}$  \cite{pdg}, and  the  Wilson coefficients 
$C_1(\mu)=1.082$,   $C_2(\mu)=-0.185$,  computed   at   $\mu=m_b$  and 
$\bar{\Lambda}_{\mathrm{MS}}=225\MeV$    \cite{buras}. 

In order to compare the branching ratio in Eq.~(\ref{result}) with the branching fraction 
obtained experimentally in Eq.~(\ref{branching}), we might use the results found in 
Ref.~\cite{Dias:2012ek}:
\begin{equation}
  \mathcal{B}(Y(4260) \to J/\psi \:\pi^+\pi^-) = (4.3 \pm 0.9)\times10^{-2}\,,
  \label{brY}
\end{equation}
and then, considering the uncertainties, we can estimate
${\mathcal B}_{_Y} >\! 3.0 \times 10^{-8}$.
However, it is important to notice that the authors in Ref.~\cite{Dias:2012ek} have considered 
two pions in the final state coming only from intermediate states, e.g. $\sigma$ and $f_0(980)$ 
mesons, which could indicate that the result in Eq.~(\ref{brY}) can be underestimated. 
In this sense, considering that the main decay channel observed for the $Y(4260)$ state is 
into $J/\psi \:\pi^+\pi^-$, we would naively expect that the branching ratio into this channel could 
also be $\mathcal{B}(Y(4260) \to J/\psi \:\pi^+\pi^-) \sim 1.0$, which would lead to the following 
result, ${\mathcal B}_{_Y} <\! 1.8 \times 10^{-6}$.
Therefore, we obtain an interval on the branching fraction 
\begin{equation}
  3.0 \times 10^{-8} < {\mathcal B}_{_Y} < 1.8 \times 10^{-6}
\end{equation}
which is in agreement with the experimental upper limit reported by Babar Collaboration 
given in Eq.~(\ref{branching}). In general the experimental evaluation of the branching fraction 
takes into account additional factors related to the numbers of reconstructed events for the 
final state ($J/\psi \:\pi^+\pi^- \:K$), for the reference process ($B \to Y(4260) \:K$), and for the 
respective reconstruction efficiencies. However, since such information has 
not been provided in Ref.~\cite{babary2}, we have neglected these factors in the calculation of 
the branching fraction ${\mathcal B}_Y$. Therefore, the comparison of our result with the 
experimental result could be affected by these differences.

In conclusion, we have used the QCDSR approach to evaluate the production of the $Y(4260)$ 
state, considered as a mixed  charmonium-tetraquark state, in  the decay 
$B\to YK$.  Using the factorization hypothesis, we find that the sum rules 
result in Eq.~(\ref{result}), is compatible with the experimental  upper limit.
Our result can be interpreted as a  lower limit for the branching ratio,
since we did not considered the  non-factorizable contributions. 
 
Our  result  was  obtained   by  considering  the  mixing  angle  in 
Eq.~(\ref{jmix}) in the range $\theta=(53.0\pm0.5)^0$. This
angle was determined in Ref.~\cite{Dias:2012ek} where the mass and
the decay width of the $Y(4260)$ in the channel $J/\psi\pi^+\pi^-$
were determined in agreement with experimental values. Therefore,
since there is no  new free  parameter in  the present  analysis,
the result   presented  here   strengthens  the   conclusion   reached  in 
\cite{Dias:2012ek} that the $Y(4260)$ is probably a mixture 
between a $c\bar{c}$ state and a tetraquark state. 

As discussed in \cite{x3872prod}, it is not simple to determine the
charmonium and the tetraquark contribution to the state described by the
current in Eq.~(\ref{jmix}). From Eq.~(\ref{jmix}) one can see that, besides 
the $\sin\theta$, the $c\bar{c}$ component of the current is multiplied by a 
dimensional parameter, the quark condensate, in order to have the same 
dimension of the tetraquark part of the current. 
Therefore, it is not clear that only the 
angle in Eq.~(\ref{jmix}) determines the percentage of each component.  
One possible way to evaluate the importance of each part of the current it is 
to analyze what one would get for the production rate with each component, 
{\it i.e.}, using $\theta=0$ and $90^\circ$ in Eq.~(\ref{jmix}). Doing this
we get respectively for the pure tetraquark and pure charmonium: 
\begin{eqnarray}
  {\mathcal{B}}(B\to Y_{\mathrm{tetra}}K) &=& (1.25\pm0.23)\times10^{-6}\,, \\
  {\mathcal{B}}(B\to Y_{\bar{c}c}K) &=&(1.14\pm0.20)\times10^{-5}\,.   
\end{eqnarray}
Comparing the results  for the pure states with the  one for the mixed 
state (\ref{result}), we can see that the branching ratio for the pure 
tetraquark is  one order smaller,  while the pure charmonium  is larger.  
From these  
results we see that the $c\bar{c}$ part of the state plays a very important 
role in the determination of the branching ratio. On the other hand, in the 
decay $Y\to J/\psi\pi^+\pi^-$, the width  
obtained in our approach for a pure $c\bar{c}$ state is \cite{Dias:2012ek}: 
\begin{equation}\lb{xppcc} 
  \Gamma(Y_{\bar{c}c}\to J/\psi\pi\pi)=0\,, 
\end{equation} 
and, therefore, the tetraquark part of the state is the only one  
that contributes to this decay, playing an essential role  
in the determination of this decay width. 
 
Therefore, although we can not determine the percentages of the $c\bar{c}$ 
and the tetraquark components in  the $Y(4260)$, we may say that both 
components are extremely important, and that, in our approach, it is not  
possible to explain all the experimental data about the $Y(4260)$ with 
only one component. 
\vspace{0.8cm}
 
\section*{Acknowledgment} 
 
\noindent  
This work has been partially supported by S\~{a}o Paulo Research Foundation 
(FAPESP), grant n.\,2012/22815-3, and National Counsel of Technological and 
Scientific Development (CNPq-Brazil). 
\vspace{0.8cm}

\appendix
\section{Spectral Densities for the Two-point Correlation Function}
\noindent  
We list the spectral densities for the invariant functions related to the
coupling between the current $J^{(\bar{c}c)}_\mu$ and the $Y(4260)$ state.
We consider the OPE contributions up to dimension-five condensates 
and keep terms at leading order in $\alpha_s$. In order to retain the 
heavy quark mass finite, we use the momentum-space expression for 
the heavy quark propagator. We calculate the light quark part of the 
correlation function in the coordinate-space and use the Schwinger 
parametrization to evaluate the heavy quark part of the correlator. For
the $d^4y$ integration in Eq.~(\ref{2point}), we use again the Schwinger 
parametrization, after a Wick rotation. Finally, the result of these integrals 
are given in terms of logarithmic functions through which we extract the 
spectral densities. The same technique can be used for evaluating the 
condensate contributions.

Then, in the $g_{\mu\nu}$ structure, we evaluate the spectral densities for 
the $\Pi^{2,2}(M_B^2)$ function,  
\begin{eqnarray}
  \rho^{2,2}(s) \!&=&\! \frac{m_c^2}{4\pi^2} ~v\Big( 2+ \frac{1}{x} \Big) +
  	\frac{\gGG}{48\pi^2} ~\frac{v}{M_B^2}\bigg[ 4\Big( 1-\frac{1}{x} \Big) \hspace{1cm} \nn\\
	&& ~- \frac{m_c^2}{M_B^2 \:x} \Big( 11 - \frac{5}{x} \Big)
	+ \Big( \frac{m_c^2}{M_B^2 \:x} \Big)^2 \Big( 3 - \frac{1}{x} \Big) \bigg],
\end{eqnarray}
and for the $\Pi^{2,4}(M_B^2)$ function,
\begin{eqnarray}
  \rho^{2,4}(s) \!&=&\! -\frac{m_c^2 \:\qq}{12\pi^2} ~v\Big( 2 \!+\! \frac{1}{x} \Big) +
  	\frac{\qGq}{24\pi^2} ~v \Big( 1 \!-\! \frac{m_c^2}{M_B^2 \:x} \Big) \hspace{0.7cm}
\end{eqnarray}
where we have used the definitions
\begin{eqnarray}
  x &=& m_c^2 /s \\
  v &=& \sqrt{1-4x} ~~.
\end{eqnarray}

\end{document}